\documentclass[fleqn,usenatbib]{mnras}

\usepackage{newtxtext,newtxmath}

\usepackage[T1]{fontenc}

\usepackage{soul}
\usepackage{xcolor}

\DeclareRobustCommand{\VAN}[3]{#2}
\let\VANthebibliography\thebibliography
\def\thebibliography{\DeclareRobustCommand{\VAN}[3]{##3}\VANthebibliography}

\usepackage{graphicx}	
\usepackage{amsmath}	

\title[Chemical evolution of a nitrogen-rich object]{Chemical evolution of a young super star cluster at the Sunburst Arc}

\author[Tapia et al.]{
Truman Tapia \thanks{E-mail: truman.tapiamora@research.uwa.edu.au},$^{1}$
Kenji Bekki,$^{1}$
Brent Groves$^{1}$\\
$^{1}$ International Centre for Radio Astronomy Research, The University of Western Australia, 35 Stirling Highway, Crawley, Western Australia 6009, Australia\\
}

\date{Accepted XXX. Received YYY; in original form ZZZ}

\pubyear{2015}

\begin{document}
\label{firstpage}
\pagerange{\pageref{firstpage}--\pageref{lastpage}}
\maketitle

\begin{abstract}
Recent observations of high-redshift galaxies have revealed starburst galaxies with excessive amounts of nitrogen, well above that expected in standard evolutionary models. The Sunburst Arc galaxy, particularly its young and massive star cluster, represents the closest ($z=2.4$) and brightest of these as a strongly lensed object. In this work, we study the chemical history of this star cluster to determine the origin of the elevated gas-phase nitrogen using a chemical evolution model. Our model includes the enrichment of OB stars through stellar winds and core-collapse supernovae assuming that massive stars ($M>25$ $M_\odot$) collapse directly into black holes at the end of their lives. We fit the model parameters to the observed chemical abundances of the Sunburst Arc cluster: O/H, C/O, and N/O. We find that the observed chemical abundances can be explained by models featuring intense star formation events, characterized by rapid gas accretion and high star formation efficiencies. Additionally, the stellar population contributing to the gas enrichment must exclude Wolf-Rayet stars. These conditions might be present in other nitrogen-rich objects as their similar chemical abundances suggest a common history. As previous studies have proposed the presence of Wolf-Rayet stars in the new nitrogen-rich objects, further research using chemodynamic modeling is necessary to ascertain the true nature of these objects.

\end{abstract}




\section{Introduction}

Recent observations reveal high-redshift objects with unusually strong nitrogen (N) emission lines \citep{pascale2023nitrogen, marques2024extreme, vink2023very, pascale2024young}, with GN-z11 being a prime example \citep{senchyna2023gn, cameron2023nitrogen}. These objects, referred to as N-emitters due to their intense N lines in the spectrum, stand out because their high N/O abundance ratios deviate from the enrichment observed in the local Universe. N-emitters are characterized by compact and intense star-forming regions \citep{2024A&A...687L..11S, pascale2023nitrogen, 2023ApJ...952...74T}, with gas-phase N/O ratios similar to the N/O stellar abundances observed in the enriched populations of local globular clusters (GCs)(see Fig.~5 in \citet{senchyna2023gn}). The gas observed in N-emitters might be locked into stars and give rise to multiple stellar populations as observed in GCs. Currently, the origin of the multiple populations in GC is insufficiently understood (see \citet{bastian2018multiple} for a review of the multiple population problem). The characteristics of N-emitters place them as progenitors of GCs, offering new insights into the unresolved evolution of these ancient star clusters.

The list of N-emitters includes the galaxy GN-z11 at $z \sim 10.6$ \citep{oesch2016remarkably, cameron2023nitrogen, 2023A&A...677A..88B}, the galaxy CEERS-1019 at $z \sim 8.7$ \citep{Zitrin_2015, tang2023jwst, 2023ApJ...953L..29L, marques2024extreme}, the galaxy SMACS2031 at $z \sim 3.5$ \citep{10.1111/j.1365-2966.2012.22007.x, 10.1093/mnras/stv2859}, and the young massive star cluster in the Sunburst Arc galaxy \citep{rivera2019gravitational, pascale2023nitrogen} at $z \sim 2.4$. This latter stands out as an ideal candidate to study N-emitting objects because it is gravitational lensed and has a lower redshift than the rest of the objects.

The Sunburst Arc galaxy was discovered by \citet{dahle2016discovery} and stands out as one of the brightest gravitationally lensed galaxies known to date. The Sunburst Arc is magnified by the foreground galaxy cluster PSZ1 G311.65-18.48 [$z\sim 0.443$] into twelve distinct images arising from a disrupted Einstein ring configuration. \citet{rivera2019gravitational} identified a star-forming region within the lensed galaxy, marked by a triple-peaked $\text{Ly}\alpha$ line and escaping Lyman continuum (LyC) radiation. These findings suggest the existence of optically thin channels sculpted into the surrounding interstellar medium (ISM).

Subsequent investigations have unveiled that the star-forming region within the Sunburst Arc constitutes a nascent super star cluster (hereafter referred to as the LyC cluster) \citep{chisholm2019constraining, mainali2022connection}. The LyC cluster exhibits a youthful age of approximately 3 million years and has a mass quantified at approximately $10^7 M_\odot$ \citep{vanzella2022high}. Additionally, the LyC cluster demonstrates compact characteristics with an effective radius of approximately $8$ pc \citep{vanzella2022high}. These estimates clearly make the LyC cluster a proto-GC candidate.

\citet{pascale2023nitrogen} estimated the physical parameters of the LyC cluster through spectral analysis of HST images, ground-based VLT/MUSE, and VLT/X-shooter spectroscopy data. The authors determined that the cluster is composed of two parts, a compact central part and an extended part of outskirt stars. To explain the observed emission line structure, they found the surrounding ISM to be composed of two parts; a high density and pressure gas within 5-10 pc of the cluster center, experiencing strong radiative forces, and lower density and pressure clouds much further from the cluster. Through photoionization modeling, \citet{pascale2023nitrogen} found that the ISM surrounding the LyC cluster is subsolar ($Z\sim 0.2-0.3$ $Z_\odot$), yet the high-pressure clouds show an enhanced nitrogen of $\log(\rm N/O) \sim -0.21$ well above that observed in HII regions of similar metallicity \citep{2017MNRAS.466.4403N, 10.1111/j.1365-2966.2006.10892.x}. Recently, the LyC cluster was observed using JWST with NIRCam and NIRSpec in Cycle 1. From the obtained spectrum, \citet{rivera2024sunburst} report a high gas-phase N/O ratio of $\log(\rm N/O) \sim -0.7$ along with a low metallicity $Z\sim (0.1-0.2)$ $Z_\odot$. Using the same spectrum, \citet{welch2024sunburst} report $\log(\rm N/O) \sim -0.7$ and a subsolar metallicity of $12+\log(\rm O/H) \sim 8$ by estimating the electron temperature and density from auroral emission lines. Such high N abundances are not observed in local HII regions of similar metallicities.

This work delves into the origin of the unusual abundances observed in the LyC cluster, both as a rare object that JWST is finding more examples of, and as a proto-GC that might give insight into globular cluster formation. Using a one-zone chemical evolution model, we study the conditions that led to the observed elemental abundances. While other studies suggest Super Massive Stars (SMSs) or Very Massive Stars (VMSs), with masses much higher than a 100 $M_\odot$, as the origin of the high N enrichment \citep{charbonnel2023n, vink2023very, Nagele_2023}; we consider a less extreme scenario where OB stars are the polluters.

\section{Modelling the formation and enrichment of the Sunburst Arc cluster}
\label{sec:model}

Chemical evolution models investigate the production and distribution of elements in a system to link the current abundance distribution with the history of the system \citep{audouze1976chemical, matteucci2012chemical}. Previously, we used such models to study various systems such as dwarf galaxies \citep{bekki2012chemical}, GCs \citep{bekki2023globular}, and more recently the galaxy GN-z11 \citep{10.1093/mnrasl/slad108}. In this study, we introduce a new model to simulate the formation and enrichment process of the LyC cluster building upon our previous work. Our new model is mainly altered in the stellar population where we incorporate the work of \citet[][hereafter LC18]{limongi2018presupernova}. While the fundamental aspects of the model remain consistent with our prior work (for additional details on the model fundamentals, see \citet{bekki2012chemical}), we provide a summary.

\subsection{Model}

\subsubsection{General equations}

The model numerically integrates a set of equations to calculate the time evolution of the chemical abundances within a nascent star cluster and surrounding gas. First, we propose that the star cluster originates from a rapid infall of gas onto the Sunburst cluster. The gas is accreted at a variable rate $A(t)$ at a time $t$ since the infall began. The accreted gas forms stars at a rate $\psi(t)$, creating the cluster. The mass of the system consists of the total accreted gas, with any pre-existing ISM in the cluster not included in our calculations. This means the total mass is simply the stars formed and the gas remaining, $M_{g}$. Note that in the following equations, we normalize the total mass at the end of the simulation and the variables are dimensionless.

The newly formed stars release stellar winds that are recycled back into the surrounding gas at an emission rate $W(t)$. These winds, and at a later stage core-collapse supernovae (CCSNe), also elementally enrich the gas, increasing the abundance $Z_i(t)$ of each element, $i$. In our model, the initial element abundances are determined solely by the CCSN enrichment of the stellar population. This assumption implies that CCSNe pre-enriched the infalling gas before the LyC cluster was born. The basic equations that describe the one-zone chemical evolution model are;

\begin{equation}
    \label{eq:gas_mass}
    \frac{d M_g}{d t}=A(t)-\psi(t)+\mathit{W}(t),
\end{equation} 
\begin{equation}
\begin{split}
&\frac{d\left(Z_i M_g\right)}{d t}=Z_{A, i} A(t)- Z_i(t) \psi(t)\\
&+ \sum_{k} y_{\mathrm{w}, i, k} \psi(t-t_{\mathrm{w}, k})  + \sum_{k} y_{\mathrm{c}, i, k} \psi(t-t_{\mathrm{c}, k}),
\end{split}
\end{equation}
where the term $Z_{A, i}$ is the abundance of the $i$th element contained in the accreted gas which is determined by the initial conditions. The terms $y_{\mathrm{w}, i, k}$ and $y_{\mathrm{c}, i, k}$ represent the yields from winds and CCSNe, respectively, of the element $i$ in the stellar mass range $k$. Given the young age of the LyC cluster, we consider only the contribution from winds and CCSNe. The time $t_{\mathrm{w}, k}$ is the time delay of the stellar winds, and $t_{\mathrm{c}, k}$ the time delay of the CCSNe (i.e. the lifetime of stars of mass $k$). We integrate the chemical abundances up to $t=10$ Myr, which we consider enough to cover the observed young age of the LyC cluster.

\subsubsection{SFR and gas accretion rate}
\label{subsec:srf_A}

We consider the SFR to be dependent on the gas mass of the system $M_g(t)$ with a constant efficiency of conversion $C_\text{sf}$
\begin{equation}
\psi(t) = C_\text{sf} M_g(t).
\label{eq:sfr}
\end{equation}
$C_\text{sf}$ is a free parameter constrained during the fitting process.

The gas accretion rate follows an exponential decay. This favors the rapid accretion of gas and the early formation of the stellar population parallel to the young age of the LyC cluster. Furthermore, a gas accretion rate given by an inverse exponential has been adopted in other numerical chemical evolution models \citep{spitoni2019galactic, chiappini1997chemical, 10.1093/mnras/stu484}, and solves the G dwarf distribution problem \citep{greener2021sdss}. The accretion rate is expressed as
\begin{equation}
\label{eq:accretion}
A(t) = C_ae^{-t/t_a},
\end{equation}
where $C_a$ is a normalization constant that sets to 1 the total gas mass accreted at the end of the integration time ($t=10$ Myr). The time $t_a$ controls the timescale of gas accretion, which is our second free parameter.

\subsubsection{IMF}

To account for the mass distribution of stars at formation, the model employs an Initial Mass Function (IMF) $\Psi$ given by 
\begin{equation}
\Psi (m_I)=M_{\mathrm{s}, 0}m_I^{-\alpha},    
\end{equation}
where $m_I$ denotes the mass stars at formation. The normalization constant, $M_{\mathrm{s}, 0}$, depends on the adopted lower and upper mass cutoffs. We set the lower mass cutoff to $0.1 M_{\odot}$. The upper mass cutoff $m_u$ is the third free parameter of our model and heavily influences the chemical abundances in the system. The IMF slope value $\alpha=2.35$ corresponds to the Salpeter IMF \citep{salpeter1955luminosity}. The slope $\alpha$ is the fourth free parameter in our model.

\subsubsection{Stellar population and yields}
\label{sec:stellar_pop}

The stellar population in our model enriches the system based on the yields calculated by LC18. They calculate the evolution and yields of O and B type stars across a grid of metallicity, stellar rotation, and mass. Given the young age of the LyC cluster, we do not consider the enrichment of less massive stars. We use the yields from the set R of LC18, wherein stars more massive than $25$ $M_\odot$ directly collapse into black holes (BHs) without the occurrence of CCSNe. Stars have two ways of enrichment: wind ejecta followed by CCSN ejecta. At each time step, a fraction of the stellar winds, $f_\text{SW}$, is retained in the system. Similarly, a fraction of the CCSNe ejecta, $f_\text{SN}$, is retained at each time step. We fix $f_\text{SW}=1$ to maximize the enrichment by winds and we assume $f_\text{SN}=0.03$ based on the values used in our previous work \citet{10.1093/mnrasl/slad108}. Nonetheless, we find that $f_\text{SN}$ is not a critical parameter to reproduce the observations.

We select stars with a metallicity [Fe/H] = -1 from the stellar models grid computed by LC18, as this value aligns with the subsolar stellar metallicities (0.5 - 0.6) $Z_\odot$ and 0.4 $Z_\odot$ reported by \citet{chisholm2019constraining} and \citet{mainali2022connection}, respectively, for the LyC cluster. Stellar rotation significantly impacts stellar evolution. LC18 consider the yields for three different rotations; 0, 150, and 300 km/s. We consider the ideal scenario where all the stellar population has the same stellar rotation $v_\text{rot}$, and we explore the extreme values of this parameter 0 and 300 km/s.  

Due to their considerable winds and chemical output, the presence of Wolf Rayet stars (WRs) in the stellar population can considerably impact a clusters chemical evolution. These stars release high concentrations of N, with studies suggesting that WRs are the main chemical polluters of systems with high N/O abundance \citep{watanabe2024empress, Kobayashi_2024}. LC18 defines that a star enters the WR phase when its effective temperature $\log(\rm T_\text{eff}/K)>4$ and its surface hydrogen abundance $H_\text{surf}<0.4$. Given that WRs have lost significant amounts of H, only the most massive stars enter this phase because of their strong stellar winds. LC18 show that the minimum initial mass for WR formation decreases as the stellar rotation increases (see Fig. 38 of LC18 for a summary of the stellar models). At [Fe/H] = -1 (the value chosen in this work), WRs form from stars with initial masses $M/M_\odot \ge$ 80, 40, and 30 for the stellar rotations 0, 150, and 300 (in km/s), respectively, in the LC18 stellar models. We study the impact of WRs for reproducing the abundances of the LyC cluster by controlling their presence in the stellar population through the upper mass cutoff of the IMF $m_u$.

\begin{table}
	\centering
	\caption{Chemical abundances of the LyC cluster used to fit our model. The values are extracted from \citet{pascale2023nitrogen}. The O/H ratio corresponds with the gas-phase metallicity of the cluster.}
	\label{tab:observed_params}
	\begin{tabular}{lr} 
		\hline
		Abundance & Value\\
		\hline
		  $\log(\text{N/O})$ & $-0.21\pm 0.1$\\
        $\log(\text{C/O})$ & $-0.51\pm 0.05$\\
		$12 + \log(\text{O/H})$ & $8.08\pm 0.09$\\
		\hline
	\end{tabular}
\end{table}

\subsection{Observations and fitting process}
\label{sec:fitting_process}

Our primary aim is to reproduce the observed chemical abundances of the nebulae surrounding the LyC cluster (i.e. the high-pressure and high-density clouds) using the chemical evolution model just described. We focus on the most abundant elements reported for the cluster; i.e. O, C, and N. Table \ref{tab:observed_params} shows the abundance ratios that our model attempts to reproduce.

Our chemical evolution model computes the time evolution of the elemental abundances based on the provided model parameters. To compare these results to observations, we define a specific time of observation. At this time, we can calculate the age of the cluster denoted by $age_\star(t)$\footnote{$age_\star$ is defined as the mass-weighted mean stellar age in our model.}. The cluster age is included as a free parameter that is constrained during the fitting process. 

Assuming the total gas mass accreted is $10^7$ $M_\odot$ (aligning with the cluster mass reported by \citet{pascale2023nitrogen}), we introduce physical units to the model. Specifically, $C_\text{sf}$ has units of inverse time (see equation \ref{eq:sfr}), such that $C_\text{sf}=1=10^{-8} \text{ yr}^{-1}$. This physical parameter can be interpreted either as the inverse depletion time or as the star formation efficiency (SFE), $\varepsilon_\text{sf}$. In the following sections, we will use the latter term when reporting the model parameters. In total, we have 5 adjustable parameters: the upper mass cutoff ($m_u$), SFE ($\varepsilon_\text{sf}$), IMF slope ($\alpha$), gas accretion timescale ($t_a$), and the cluster age ($age_\star$). Table \ref{tab:free_params} shows these free parameters as columns.

\begin{table*}
	\centering
	\caption{The parameters of the best-fitting models. Columns in bold represent the free parameters with their units and range of exploration during the fitting process. From left to right the parameters are the upper-mass cutoff, star-formation efficiency, IMF slope, gas accretion timescale, and cluster age. The no rotation models have a stellar population without stellar rotation ($v_\text{rot}=0$), whereas the rotation model has stars rotating at $v_\text{rot}=300$ km/s. The uncertainties of the best-fitting models include the 68\% confidence interval of the posterior distributions obtained through MCMC sampling.}
	\label{tab:free_params}
	\begin{tabular}{lccccr} 
		\hline\hline
         &  $\boldsymbol{m_u}$  &  $\boldsymbol{\varepsilon}_\textbf{sf}$  &  $\boldsymbol{\alpha}$  &  $\boldsymbol{t_a}$  &  $\boldsymbol{age_\star}$\\
         Best-fitting Models &  [$M_\odot$]  &  [$10^{-7}$ yr$^{-1}$]  &  &  [Myr]  &  [Myr]\\
         &  (15 - 120)  &  (1 - 15)  &  (1.6 - 2.8)  &  (0.5 - 5)  &  (2 - 6)
         \\
        \hline
        No rotation  & $80.9_{-10.3}^{+0.5}$  &  $11_{-1.8}^{+2.9}$  &  $1.9\pm 0.2$  &  $0.8_{-0.2}^{+0.3}$  &  $4.8_{-0.5}^{+0.6}$\\ \\
        Rotation  &  $30.7_{-0.}^{+1.1}$  &  $11.5_{-5.5}^{+1.4}$  &   $1.7_{-0.}^{+0.5}$  &  $2.1_{-1.1}^{+0.2}$  &  $5.4_{-0.3}^{+0.4}$\\ \\
        Salpeter no rotation  & 81.3  &  15  & 2.35  &  0.8  &  4.6\\
		\hline
	\end{tabular}
\end{table*}

We use least squares to fit the model to the observations and obtain the posterior distributions of the free parameters. We use the routine \textit{minimize} from \textit{scipy} \citep{2020SciPy-NMeth} to find the best-fitting model with an optimizer. Additionally, we explore the parameter space using \textit{emcee} \citep{foreman2013emcee}, which runs a Markov Chain Monte Carlo (MCMC) simulation to sample the probability distribution of the free parameters. We take the 68\% confidence intervals of the posterior distributions to report the uncertainties. The distributions of the parameters obtained through MCMC clarify which physical scenarios can reproduce the observations and which cannot. We execute the MCMC simulation with 10 different initial conditions for 4000 iterations. 

\begin{figure*}
	\includegraphics[width= \textwidth]{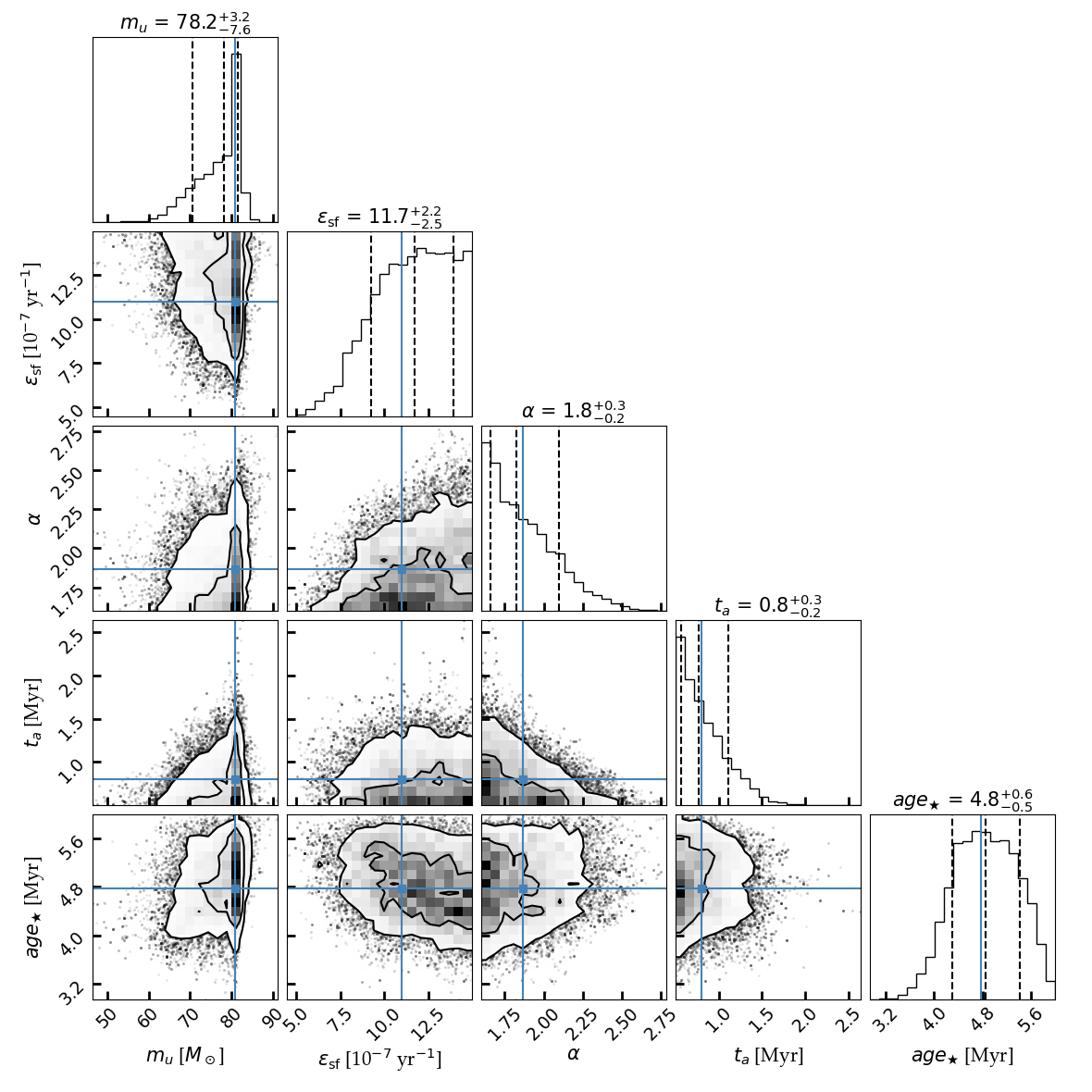}
    \caption{Posterior distributions for the free parameters studied in this work obtained from MCMC sampling. The parameters correspond to a simulated star cluster with non-rotating stars. 1D histograms show the marginalized posterior probability distributions of the parameters. These are marked with the 68\% confidence intervals in dashed lines. Numbers with uncertainties report this interval around the median. 2D contours show the correlation of the posterior distributions with the contours highlighting the 50\% and 90\% density levels. Light blue lines signal the best-fitting model. The values of this model as well as the free parameters description are shown in Table \ref{tab:free_params}.}
    \label{fig:mcmc}
\end{figure*}

\section{Results}
\label{sec:results}

\begin{figure}
	\includegraphics[width= \columnwidth]{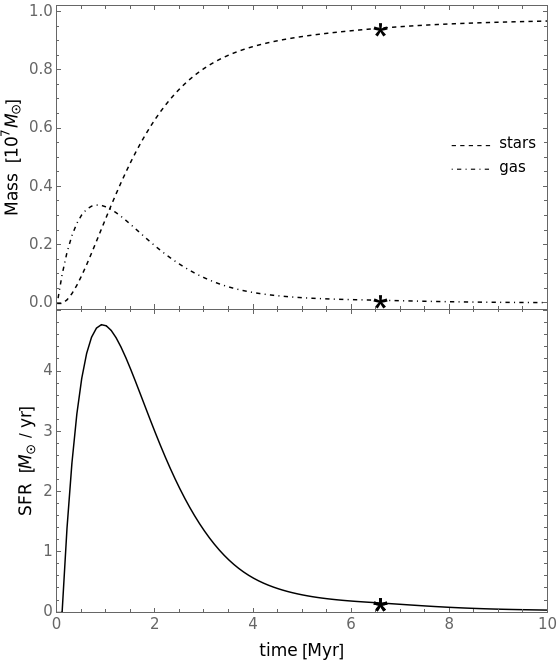}
    \caption{Top panel: time evolution of the stellar mass (dashed curve) and the gas mass (dot-dashed curve) of the best-fitting model without stellar rotation. Bottom panel: time evolution of the SFR of the same model. The asterisks signal the time when the model fits to the observables, namely the present cluster state. The model parameters are displayed in Table \ref{tab:free_params}.}
    \label{fig:fiducial_mass}
\end{figure}

\begin{figure}
    \includegraphics[width= \columnwidth]{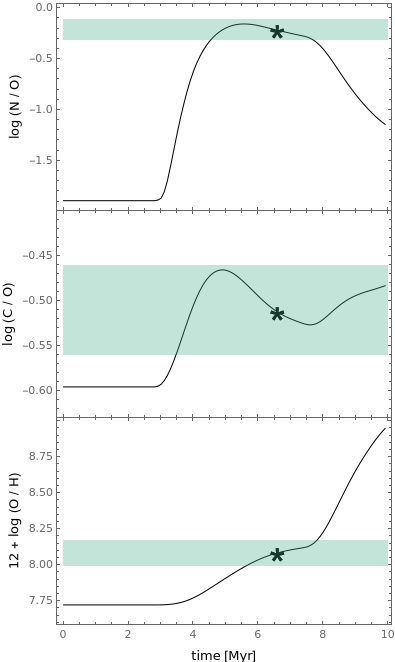}
    \caption{Time evolution of the abundances $12+\log(\rm O/H)$ (bottom), $\log(\rm C/O)$ (central), and $\log(\rm N/O)$ (top) of the best-fitting model. Shaded areas mark the chemical abundances and uncertainties reported by \citet{pascale2023nitrogen} for the LyC cluster. The asterisks signal the time when the model fits to the observations, namely the present cluster state. The model parameters are described in Table \ref{tab:free_params}.}    
    \label{fig:fiducial_abundances}
\end{figure}

\subsection{Best-fitting model (no stellar rotation)}

\label{sec:best}

Using the aforementioned model and fitting procedure, we first limit our exploration to non-rotating stars and explore the impact of rotation in Section \ref{subsec:best_rot}. Table \ref{tab:free_params} presents the values of the best-fitting model (referred to as the No rotation model), with the parameter space exploration presented in Fig. \ref{fig:mcmc} (best-fitting values are shown in light blue). The time evolution of the best-fitting model is shown in Figs. \ref{fig:fiducial_mass} and \ref{fig:fiducial_abundances}. The asterisk marks the observing time, with the associated cluster age value reported, at which the model reproduces observed abundances (shown as the green bands). Notice that the cluster age is defined as the mass-weighted mean stellar age (see Section \ref{sec:fitting_process}) and the time $t=0$ marks the start of gas accretion. The found cluster age is $4.8_{-0.5}^{+0.6}$ Myr. Compared with the results of \citet{pascale2023nitrogen}, our cluster age is higher than their fiducial age of 2.43 Myr, but our estimate overlaps with their age estimates for top-heavy and continuous star formation scenarios ($3.3^{+1.2}_{-1.8}$ and $4.3^{+1.7}_{-1.9}$ Myr, respectively) that better match our model characteristics and assumptions. Also, our cluster age overlaps with the age reported by \citet{rivera2024sunburst} of (4.2 - 4.5) Myr. 

The best-fitting IMF slope is flatter compared to Salpeter at $\alpha = 1.9$. Similarly, our previous study of another N-emitter, GN-z11, found that a top-heavy IMF was required to explain the observations \citep{10.1093/mnrasl/slad108}. The gas accretion timescale of the cluster is $0.8_{-0.2}^{+0.3}$ Myr in our best model; hence, the cluster is formed from a fast ingestion of high amounts of gas. We find a high value of SFE $\varepsilon_\text{sf}\sim 1.1\times10^{-6}\text{yr}^{-1}$ compared to the typical efficiencies observed in local spiral galaxies of $\varepsilon_\text{sf}\lesssim 10^{-8}\text{yr}^{-1}$ \citep{leroy2008star, hagedorn2024molecular}. This high SFE is required because it leads to a rapid formation of stars, enabling the early enrichment. The upper mass cutoff has a value of $m_u=80.9_{-10.3}^{+0.5}$, which constrains the stellar population responsible for the enrichment. Specifically, the upper bound of this value coincides with the WR formation threshold found in LC18 for non-rotating stars at [Fe/H] = -1 (see section \ref{sec:stellar_pop}). Hence, we find that WR stars (as modeled within LC18) must not be present to explain the observed gas-phase abundances in the LyC cluster. Furthermore, our calculated value for the upper mass cutoff implies that Blue Super Giants (BSGs) are the main polluters of the system (for a summary of the types of stars for different initial masses, refer to Fig. 38 of LC18). 

Fig.~\ref{fig:fiducial_mass} (top panel) shows the time evolution of the mass of the best-fitting model. The model begins with a rapid infall of gas because of the short gas accretion timescale $t_a=0.8$. Before $t=t_a$, the gas mass is higher than the stellar mass because the gas accretion dominates over the transformation of this gas into stars. Later, the gas mass decreases as the SFR begins to dominate over the declining accretion rate. Such transformation of gas into stars steeply increases the stellar mass from t = 0 until 4 Myr when the rate of star formation is small because little gas is left in the system. Additionally, the stellar winds contribute to the gas mass because they are recycled in our model. This contribution of $\lesssim 0.1$ $M_\odot/\text{yr}$ is small and does not change the mass content of the system, but it changes the gas-phase elemental abundance ratios. Overall, the short accretion timescale allows the quick formation of the stellar population from the massive gas cloud.

In Fig.~\ref{fig:fiducial_mass} (top panel), the gas mass is a small fraction of the stellar mass at the observing time signaled by the asterisk. Specifically, $M_g\sim 1.1\times10^5$ $M\odot$, while the stellar mass $M_\star\sim 9.7\times10^6$ $M_\odot$ at this time. In agreement with this result, \citet{pascale2023nitrogen} estimate the mass of the high-pressure clouds surrounding the cluster to be $\sim 10^5$ $M_\odot$. Moreover, we obtain a gas-to-star conversion efficiency near unity considering that almost all the initial infalling gas is converted into stars in our model. Concerning this, \citet{marques2024witnessing} reports an efficiency for the LyC cluster $\gtrsim 0.45$ from simple energy balance calculations. Yet, from their Fig. 8 (right panel), the efficiency has to be $\gtrsim 0.9$ when only the feedback of stellar winds is considered, which aligns with our assumptions and findings. Additionally, the low gas mass obtained in our model would favor the escape of Ly-continuum radiation from the cluster, which matches with the high escape fraction of this radiation observed in the cluster \citep{rivera2019gravitational}.

Fig.~\ref{fig:fiducial_mass} (lower panel) shows the time evolution of the SFR of the best-fitting model. The SFR mirrors the gas mass as they are proportional by definition (see equation \ref{eq:sfr}). The high SFE enhances the SFR early on. This produces the large number of stars required to pollute the gas and reproduce the high levels of enrichment observed in the cluster. The mean SFR during the assembly of the cluster ($t\le4$ Myr) is $\sim2.5$ $M_\odot/\text{yr}$. The SFR calculated by \citet{pascale2023nitrogen} in their constant star formation scenario depends on the magnification factor of the cluster $\mu_9$. Assuming that $\mu_9=30$ (a plausible magnification value) their $\text{SFR}\sim4.5$ $M_\odot/\text{yr}$ is higher than our value. Most of the star formation has ceased at the time of observation (signaled by the asterisks in Fig.~\ref{fig:fiducial_mass}) in the modeled cluster.

Fig.~\ref{fig:fiducial_abundances} shows the chemical abundances of the best-fitting model as a function of time. The model successfully reproduces the gas-phase abundance ratios of O/H, C/O, and N/O measured in the LyC cluster. The chemical evolution of the modeled cluster can be divided into three stages: (1) pre-enrichment stage, (2) winds stage, and (3) winds + CCSNe stage. In the first stage, the constant abundances reflect those of the infalling gas. The winds stage of enrichment starts at approximately $t = 3.5$ Myr when the abundances increase for the first time driven by the winds of BSGs. These stars are the most massive of the stellar population with the shortest lifetimes. The drastic N/O enrichment observed at this stage can be explained by the high SFR and the shallow IMF slope of the best-fitting model. The direct collapse of massive stars into BHs prolongs the winds stage, as the first CCSN occurs much later, allowing greater wind-only enrichment. It is during this wind enrichment phase that the chemical abundances of the best-fitting model reproduce the observations. The last enrichment stage starts at approximately $t = 8$ Myr when the first CCSNe occur in the cluster coming from stars with a mass $M=25$ $M_\odot$. Along this stage, an increasing number of stars enrich the system through CCSNe, while the enrichment coming from winds decreases as the most massive stars end their lives. CCSN ejecta add O to the gas rapidly enriching it and decreasing the N/O abundance ratio below the value measured in the LyC cluster.

\subsection{Stellar rotation}

\label{subsec:best_rot}

\begin{figure}
	\includegraphics[width= \columnwidth]{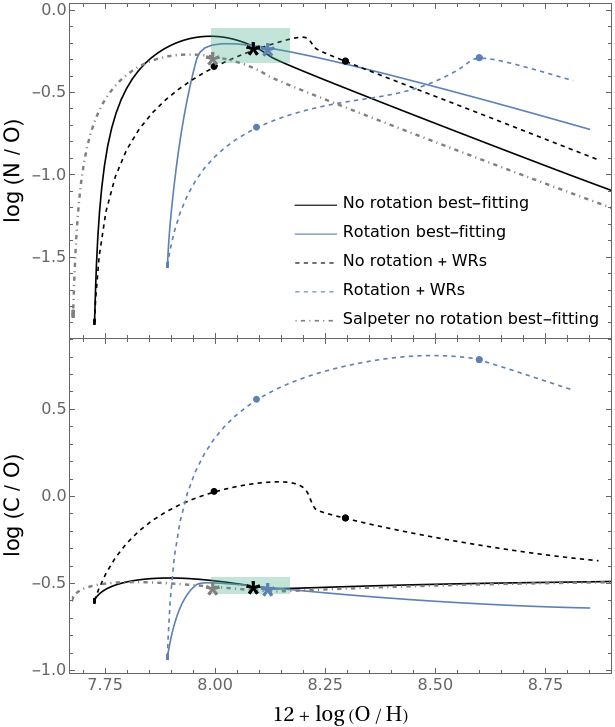}
    \caption{Chemical evolution of log(N/O) (top panel) and log(C/O) (bottom panel) versus 12 + log(O/H). Shaded regions and asterisks denote the same elements as in Fig.~\ref{fig:fiducial_abundances}. Solid curves correspond to the best-fitting models with no rotation (black), rotation at $v_\text{rot}=300$ km/s (blue), and Salpeter no rotation (thin gray), with the model parameters detailed in Table \ref{tab:free_params}. The best-fitting cluster ages correspond to the integration times $t=6.6$ Myr for the no rotation model and $t=8.1$ Myr for the one with rotation. Dashed curves correspond to models that include WR stars ($m_u=120$) for the two extremes of the stellar rotation. The remaining parameters of these models equal their best-fitting counterparts. Dots on the dashed curves indicate integration times of 4 and 8 Myr.}
    \label{fig:bests abundances}
\end{figure}

We repeat the fitting process but with the LC18 models that have a stellar rotation of 300 km/s. Table \ref{tab:free_params} presents the values of the best-fitting model with stellar rotation (referred to as the Rotation model), with the uncertainties derived from MCMC sampling. This model can reproduce the observed abundances as shown in Fig.~\ref{fig:bests abundances}. The rotation model has a low upper mass cutoff of $m_u=30.65$ $M_\odot$. This value aligns with the WR formation threshold at a stellar rotation of 300 km/s in the LC18 models (see Section \ref{sec:stellar_pop}), demonstrating that WR enrichment must be avoided as in the case with no rotation. Furthermore, the rotation best-fitting model has a high SFE and a shallow IMF slope, which aligns with the non-rotating case. The gas accretion timescale and the cluster age of the two best-fitting models (with and without stellar rotation) show differences. In the rotating case, the accretion timescale is greater which implies an extended and less abrupt accretion event. Also, the cluster shows an older best-fitting cluster age when stellar rotation is included. This can be explained by the lower mass of the stellar population.

Fig.~\ref{fig:bests abundances} shows the chemical evolution of the best-fitting models (in solid lines) at the two extremes of the stellar rotation. Given that the initial conditions are set based on the CCSN enrichment of each stellar population, the models have different initial abundance ratios (the evolution starts at the bottom left). A winds-only phase drives the steep increase of the N/O ratio in the models. Later, the CCSNe start along with the decrease of the N/O ratio. The two models reproduce the observed chemical abundances, but at different enrichment stages, with the rotating model reproducing them at $t=8.1$ Myr when CCSNe are already ongoing. 

\subsection{Unfeasible physical scenarios}

The best-fitting models show that WR enrichment must be avoided to reproduce the observations regardless of the stellar rotation. In Fig.~\ref{fig:bests abundances}, we present models with WR enrichment in dashed curves. Although enrichment from WR winds leads to a high N/O ratio, it also leads to a high C/O ratio inconsistent with the value reported for the LyC cluster (signaled by the shaded region). Hence, the C/O observation challenges the models where WRs (as modeled in LC18) enrich the gas of the cluster.

\begin{figure}
	\includegraphics[width=\columnwidth]{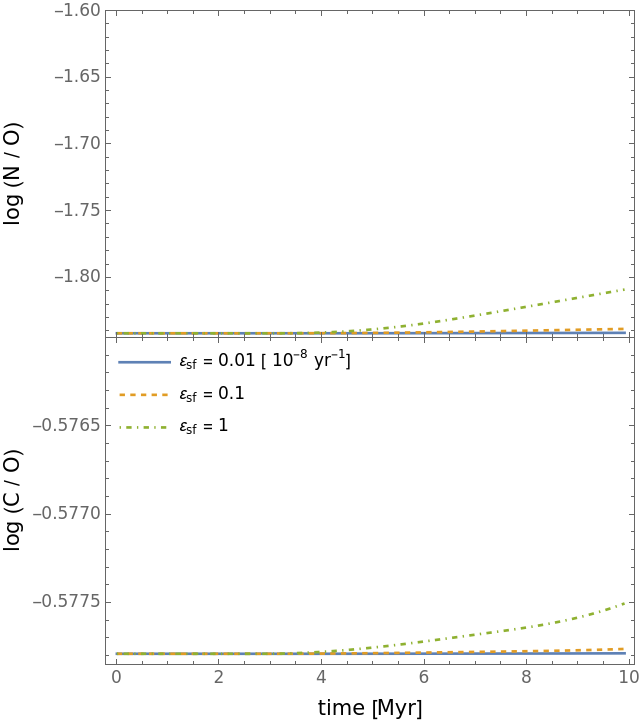}
    \caption{Chemical evolution of log(N/O) (top panel) and log(C/O) (bottom panel) versus 12 + log(O/H). Shown are models with low SFEs as indicated in the legend. All models assume non-rotating stars, with parameters identical to the best-fitting model (see Table \ref{tab:free_params}) except for $\varepsilon_\text{sf}$.}
    \label{fig:evolution_galaxies}
\end{figure}

In the posterior distributions of the SFE shown in Fig.~\ref{fig:mcmc}, low efficiencies ($\varepsilon_\text{sf}<5\times 10^{-7}\text{yr}^{-1}$) are not part of the distribution. This implies that models with low SFEs, as those observed in galaxies in the Local Universe ($\lesssim 10^{-8}\text{yr}^{-1}$) \citep{leroy2008star, hagedorn2024molecular}, cannot reproduce the observations. To exemplify this scenario, we show in Fig.~\ref{fig:evolution_galaxies} the chemical evolution of models with such low SFEs. For these models, the C/O and N/O abundance ratios scarcely increase during the 10 years of integration time, which is incompatible with the high N enrichment observed in the LyC cluster.

\subsection{Parameter degeneracies}

\label{subsec:degeneracies}

The 2D histogram of the IMF slope ($\alpha$) and the gas accretion timescale ($t_a$) in Fig.~\ref{fig:mcmc} shows degeneracies between these two parameters. A large number of massive stars is needed to reproduce the observed high N/O ratio, which can be achieved with different combinations of $t_a$ and $\alpha$ values. One extreme of this degeneracy lies at the high end of $\alpha$ where the IMF is more steep. In this extreme, a greater fraction of the stellar mass resides in low-mass stars, reducing the number of massive stars. To compensate for this reduction, $t_a$ takes low values, allowing gas to accumulate quickly, thereby increasing the SFR (see equation \ref{eq:sfr}) and producing a larger stellar population. The other extreme of the degeneracy occurs at the lower values of $\alpha$, where the IMF turns shallow. Here, a greater fraction of the stellar mass resides in massive stars, allowing longer accretion timescales with corresponding smaller stellar populations. In all cases, the models require a short gas accretion timescale ($t_a<1.5$ Myr) to have a sufficiently large stellar population and chemical enrichment.   

In Fig.~\ref{fig:mcmc}, the 2D histogram of $\alpha$ versus the SFE ($\varepsilon_\text{sf}$) shows another degeneracy. One of the extremes of the degeneracy occurs at the low-efficiency values. In this case, smaller stellar populations are formed initially, and to increase the number of massive stars, the IMF slope turns flatter. The other extreme of the degeneracy occurs at the high SFEs. Here, larger stellar populations are formed in the first million years, and as a result, the IMF can have a steeper slope. Despite this degeneracy, the SFE takes high values ($\varepsilon_\text{sf}>5\times 10^{-7} \text{yr}^{-1}$) in all cases.  

\subsection{Top-heavy IMF?}

The two best-fitting models discussed, one with stellar rotation and one without, both exhibit shallow IMF slopes when compared with the Salpeter value ($\alpha<2.35$). However, the broad posterior distributions obtained for $\alpha$ and its degeneracy with other model parameters allow for a model with a Salpeter IMF slope that can reproduce the observations. To illustrate this, we search for a best-fitting model fixing $\alpha = 2.35$ and assuming no stellar rotation. Fig.~\ref{fig:bests abundances} shows the chemical evolution of this Salpeter no rotation model (in the thin gray line) with the model parameters shown in Table~\ref{tab:free_params}. Although this model is less accurate in reproducing the chemical abundances of the LyC cluster, it does so within the reported uncertainties at a cluster age of 4.6 Myr, as signaled by the asterisk. Notably, the model has the highest possible value of the SFE ($\varepsilon_\text{sf}=1.5\times 10^{-6} \text{yr}^{-1}$).

\section{Discussion}
\label{sec:discussion}

\subsection{Proposed model}

This study demonstrates that the abundance ratios of C/O, N/O, and O/H of the gas around the LyC cluster, as reported by \citet{pascale2023nitrogen}, can be reproduced by a chemical evolution model in which OB stars are the main polluters. During our fitting process, we use MCMC to sample extensively the space of free parameters of our model, thereby finding accurate constraints in the physical scenarios that reproduce the observations. The summary of the best-fitting models is shown in Table \ref{tab:free_params}.

We emphasize the similarities between our two best-fitting models, with and without stellar rotation, and other studies on the LyC cluster and N-emitting objects. First, the mean SFRs of our models during cluster formation are $\sim$2.5 $M_\odot$/yr and $\sim$1.6 $M_\odot$/yr for the no-rotation and rotation cases, respectively. These values are lower but comparable to the $\sim$4.5 $M_\odot$/yr derived by \citet{pascale2023nitrogen} for their continuous star formation scenario, which is influenced by an uncertain magnification factor. Second, our gas-to-stellar mass ratio of $\sim 0.01$ agrees with their estimated ratio. Third, the mass-weighted cluster age we report for the no-rotation model, $4.8_{-0.5}^{+0.6}$ Myr, overlaps with age estimates from other studies \citep{rivera2024sunburst, pascale2023nitrogen}, while our age estimate including stellar rotation is older and shows less agreement. Fourth, the shallow IMF slope of the best-fitting models ($\alpha<2.35$) aligns with our study on GN-z11 \citep{10.1093/mnrasl/slad108}, which also exhibits a high N/O ratio. However, we demonstrated that a Salpeter slope can also reproduce the observations.

In our chemical evolution model, the direct collapse of massive stars ($>25$ $M\odot$) into BHs is a key factor for achieving a high N/O abundance ratio. This direct collapse delays the onset of CCSNe, allowing stellar winds to pollute the system for a longer period. As described in Sections \ref{sec:best} and \ref{subsec:best_rot}, this drastic wind-only enrichment increases the N/O ratio to the levels observed in the LyC cluster. Conversely, once CCSNe commence in our best-fitting models, the N/O ratio decreases. Hence, in a model where massive stars undergo CCSN, achieving high N/O ratios is not possible because a winds-only stage of enrichment cannot be obtained.

\subsection{N-emitters and other hypotheses}

As noted in the Introduction, the LyC cluster is one of several known objects with high gas-phase N/O ratios (N-emitters). Among these objects, the galaxies CEERS-1019 and SMACS2031 show moderate C/O levels, similar to those found in the LyC cluster \citep{marques2024extreme, 10.1111/j.1365-2966.2012.22007.x}. Additionally, the lower limit of the C/O ratio measured in GN-z11 shows a comparable level of enrichment \citep{cameron2023nitrogen}. These common elemental abundances, along with the presence of compact and intense star-forming regions, suggest these N-emitters have a similar chemical evolution. Furthermore, studies have suggested that these objects are proto-GCs \citep{marques2024extreme, cameron2023nitrogen, senchyna2023gn, charbonnel2023n}. If these objects have related histories, the polluters found in this work, OB stars with initial masses lower than WRs, might be the polluters acting in N-emitters. 

The origins of the chemical abundances of N-emitters remain a subject of active debate. Super Massive Stars (SMSs), with masses $M \ge 1000$ $M_\odot$, have been proposed to explain this phenomenon  \citep{charbonnel2023n, marques2024extreme}. However, the existence of SMSs is questionable as they have not yet been observed in nature. Another potential source of enrichment is Very Massive Stars (VMSs) with masses $M \ge 100$ $M_\odot$ \citep{vink2023very}. VMSs have been observed in regions such as the Tarantula Nebula in the LMC \citep{shenar2023constraints}, but their formation and evolution remain uncertain, highlighting the need for further research to support this hypothesis. \citet{mevstric2023clues} analyzed the VLT/MUSE spectrum of the LyC cluster to conclude that VMSs are present based on their spectral signatures. 

Several models have been proposed to explain the chemical abundances observed in GN-z11, but their applicability to N-emitters like the LyC cluster remains questionable. \citet{Kobayashi_2024} introduce a two-starburst model that takes WRs as part of the stellar population to explain the chemical abundances in GN-z11. Their model predicts that WR stars lead to high C/O and N/O abundance ratios, consistent with our findings. Hence, this hypothesis is unlikely for N-emitters with moderate to low C/O ratios as the LyC cluster. \citet{watanabe2024empress} model the chemical evolution of GN-z11 with the same stellar yields as our work, finding that WR winds can explain the high N/O ratio. However, they do not study the C/O ratio. \citet{d2023gn} suggest that the observed abundances in GN-z11 are due to the pollution from asymptotic giant branch stars. Nonetheless, the timescale for this scenario (several tens of million years) seems inconsistent with the young age of the LyC cluster.

Recently, \citet{rivera2024sunburst} concluded that WR stars inhabit the LyC cluster based on the detection of WR blue and red optical bumps in the cluster spectrum obtained with the JWST \citep{martins2023inferring}. This raises the possibility that WR stars contribute to the gas enrichment of the system, potentially contradicting the findings of this work. However, the WR bumps likely originate in the stellar atmospheres and do not necessarily confirm enrichment contribution. Our model remains valid within the context of these new findings if we assume that WR stars inhabit the cluster but that their winds are too energetic to remain within it, thus not contributing to the enrichment. Such kinetically energetic winds could have carved out the surrounding gas, leading to the high escape of Ly-continium radiation observed in the star cluster \citep{rivera2019gravitational}. To clarify the role of WR stars in the LyC cluster, further studies that include the dynamics of stellar winds and their host cluster are required.

\subsection{Model limitations and future directions}

Chemodynamical simulations would enhance our comprehension of proto-GCs like the LyC cluster. The polluters proposed to explain the chemistry of these GC precursors, namely the OB stars identified in this work and more massive stars such as WRs, VMSs, and SMSs, have wind velocities $\gtrsim 1000$ km/s, which are above the escape velocities of a massive star cluster. For this reason, \citet{vink2023very} argued that WR winds are too fast to be retained in a star cluster. However, the interaction of the stellar winds with the gas present in the cluster and the action of radiative cooling can retain these winds within the cluster. For instance, \citet{szecsi2019role} simulated the early formation of a star cluster including its dynamics and they found that the stellar winds can be retained in the cluster in a time window before the first 4 Myr of evolution. More dynamic simulations are needed to clarify if the stellar winds of massive stars can be retained in massive star clusters. In this regard, our model assumes that the winds of OB stars, slower than the winds of the rest of the proposed polluters, remain in the system. 

\section{Conclusion}

We have presented a chemical evolution model that reproduces the elemental abundances observed in the Sunburst Arc cluster. Our model includes the enrichment of OB stars assuming that the most massive stars with $M>25$ $M_\odot$ collapse directly into BHs without an SN onset at the end of their lives. We use a minimization routine and MCMC sampling to accurately constrain the free parameters of our model. The best-fitting models obtained give us a picture of the history and formation of the star cluster that led to its current state. Our main results are the following:
\begin{enumerate}

\item The two best-fitting modeled clusters, one with a stellar population of no rotating stars and one with stars rotating at $v_\text{rot}=300$ km/s, reproduce the observations at the young cluster ages of 4.8 and 5.4 Myr, respectively. Our youngest age estimate agrees with the estimates of other studies.

\item The non-rotating best-fitting model has a gas-to-stellar mass ratio of $\sim 0.01$ consistent with the value reported by \citep{pascale2023nitrogen}.

\item The best-fitting models require that Wolf-Rayet stars are not present in their stellar populations. We show that the enrichment of WRs (as modeled in \citet{limongi2018presupernova}) leads to a high C/O ratio incompatible with the observations.

\item Both rotation and no rotation best-fitting models have a flatter IMF slope than the Salpeter value ($\alpha<2.35$). However, we also find a best-fitting model with a fixed Salpeter IMF slope that can reproduce the observations.

\item The best-fitting models, with and without rotation, have high star formation efficiencies exceeding those observed in local galaxies. We show that models with the efficiencies of local galaxies do not significantly enrich the cluster gas.

\item The non-rotating best-fitting model features a shorter and more abrupt gas accretion event with an accretion timescale of $t_a=0.8$ Myr, compared to the rotating model, which has a timescale of $t_a=2.1$ Myr.

\end{enumerate}

The study of high-z nitrogen-rich objects like the LyC cluster is a promising area of research. These objects seem to be connected to the nitrogen-rich populations observed in GCs, whose origin is still debated. The number of these objects is still very scarce with more expected to be discovered soon by the JWST. Our work presents a model that explains the chemical abundances observed in an N-emitter. If the histories of N-emitters are similar, as suggested by their similar chemical content, the polluters found in this work can be present in the other objects as well.

\section*{ACKNOWLEDGEMENTS}

We are grateful to the anonymous referee for their insightful feedback on this manuscript.

\section*{Data availability}

The data used in this paper, including the chemical evolution models and the minimization fitting products, will be shared on reasonable request to the corresponding author.



\bibliographystyle{mnras}
\bibliography{main} 




\appendix


\bsp	
\label{lastpage}
\end{document}